RESEARCH ARTICLE

# Human-Centric and Integrative Lighting Asset Management in Public Libraries: Qualitative Insights and Challenges From a Swedish Field Study

JING LIN[1,2], (Senior Member, IEEE), PER OLOF HEDEKVIST[3], (Senior Member, IEEE),
NINA MYLLY[3], MATH BOLLEN[4], (Fellow, IEEE), JINGCHUN SHEN[5],
JIAWEI XIONG[1,6], (Graduate Student Member, IEEE),
AND CHRISTOFER SILFVENIUS[7], (Senior Member, IEEE)

[1]Division of Operation and Maintenance, Luleå University of Technology, 97187 Luleå, Sweden
[2]Division of Product Realization, Mälardalen University, 63220 Eskilstuna, Sweden
[3]Division of Measurement Science and Technologies, RISE Research Institutes of Sweden, 50462 Borås, Sweden
[4]Division of Energy Science, Luleå University of Technology, 93162 Skellefteå, Sweden
[5]Division of Construction Technology, Dalarna University, 79188 Falun, Sweden
[6]Division of Management Science and Engineering, Nanjing University of Science and Technology, Nanjing 210094, China
[7]SCANIA Technical Center, SCANIA AB, 15132 Södertälje, Sweden

Corresponding author: Jing Lin (janet.lin@ltu.se)

This work was supported by Swedish Energy Agency through the Project "Integrated Lighting Asset Management in Public Libraries through Digital Twins (Integrerad Tillgångsförvaltning för Belysning i Allmänna Bibliotek Genom Digital Tvilling)" under Project P2022-00277.

**ABSTRACT** Traditional reliability evaluation of lighting sources often assesses only 50% of a lamp's volume, which can lead to performance disparities and misapplications due to their limited reflection of real-world scenarios. To address the limitations, it is essential to adopt advanced asset management approaches that enhance awareness and provide a more comprehensive evaluation framework. This paper delves into the nuances of human-centric and integrative lighting asset management in Swedish public libraries, employing a qualitative field study to ascertain the alignment of current practices with these advanced lighting principles. Expanding library services to 20 high-latitude locations (>55° N) in Sweden, our research employed field observations, stakeholder interviews, and questionnaires, coupled with a thorough gap analysis, to understand the current landscape and stakeholder perceptions. Our findings reveal a dichotomy between the existing conditions of library lighting and the stakeholders' experiences and expectations. Despite the intention to create conducive environments, there is a clear disconnect, with overt problems and covert challenges affecting user satisfaction and efficacy of lighting management. Managers, staff, and users reported varied concerns, including eye strain and discomfort, indicative of substantial room for improvement. The study advocates for a paradigm shift in not only lighting asset management but also reliability evaluation of lighting sources, moving toward continuous improvement, and enhanced awareness and training on human-centric and integrative lighting principles.

**INDEX TERMS** Human-centric lighting, integrative lighting, lighting asset management, library buildings, reliability of lighting sources, visual and non-visual performance.

The associate editor coordinating the review of this manuscript and approving it for publication was Haidong Shao.

## I. INTRODUCTION
Lighting, once predominantly functional in architecture, has transformed into a harmonious integration of aesthetics, energy efficiency, as well as user health and well-being.

"Human-centric" [1], [2] or the more broadly accepted term, "integrative lighting" [3], is tailored to bolster both physiological and psychological effects on individuals, encompassing both visual attributes such as light intensity, flicker, glare, and colour temperature, as well as non-visual benefits like enhancing circadian rhythm and alertness [4], [5]. As delineated by the Global Lighting Industry's strategic roadmap, the ambition of fully actualizing human-centric lighting is projected for 2040 [1].

In the realm of infrastructure, the term "asset" typifies systems or individual equipment, whereas "asset management" signifies an organization's orchestrated efforts to derive value from these assets throughout their lifecycle [6], [7], [8]. Lighting Asset Management is crafted to optimize the performance, energy efficiency, and durability of lighting assets, aligning them with the specific demands of an organization. Deviating from conventional asset management, which typically foregrounds functionality, cost, and lifespan, the human-centric and integrative approach pivots around human well-being, aiming for optimal lighting conditions that suit varied human activities.

To address the limitations identified in traditional reliability assessments of lighting sources under current standards [9], it is essential to adopt advanced asset management approaches that enhance awareness and provide a more comprehensive evaluation framework.

Today, public libraries are transitioning from mere repositories of books to hubs of learning. Indoor lighting quality in such venues doesn't just influence physical health, like vision, but also factors into mental well-being, influencing moods and productivity [9], [10]. This begs the question: What is the present state of human-centric and integrative lighting in these establishments? While libraries are bustling with activity, the emphasis on indoor lighting post-construction, whether in older or newer buildings, often diminishes. This further raises the question: Have lighting asset management practices evolved in sync with advancements like LEDs? What's the stakeholders' experience and expectations regarding the human-centric and integrative lighting in public libraries?

To navigate these inquiries, we embarked on a study across 20 diverse public libraries in Sweden, ranging from Luleå in the northern coast to the cultural heart of Malmö in the south, spanning October 2022 to July 2023. Through a combination of field investigations, interviews, and questionnaires, buttressed by an in-depth gap analysis, we unearthed qualitative insights on overtly visible problems, covert challenges, and a critical juxtaposition of optimal lighting aspirations versus their real-world implementations.

After this introductory section, our paper delves into its methodology, spotlighting the qualitative approach and encapsulating a literature review on human-centric lighting, asset management, and pertinent standards. Subsequent sections unveil insights from our on-the-ground investigations and interviews, underlining observed practices, challenges, and user feedback. Later sections probe deeper into the identified challenges and proffer an extensive discourse on opportunities, recommendations, and broader implications for Swedish public libraries. Conclusively, we encapsulate the study's pivotal revelations, pointing towards future research avenues and suggestions.

## II. METHODOLOGY

Section two outlines the qualitative research methodology used to investigate human-centric and integrative lighting in Swedish public libraries. It begins by explaining the choice of this method for its depth of insight (2.1), followed by a literature review to contextualize the study (2.2). Criteria for selecting 20 varied libraries are discussed (2.3), ensuring a representative sample. Structured interviews with library stakeholders are detailed (2.4), aimed at understanding the impact and issues of existing lighting. This approach sets the foundation for the study's deeper analysis in later sections.

### A. QUALITATIVE APPROACH AND JUSTIFICATION IN THIS STUDY

To elucidate the current landscape of human-centric and integrative lighting asset management in public libraries, we adopted a qualitative research methodology encompassing the following avenues:

#### 1) STATE-OF-THE-ART REVIEW

dive into the existing literature helped us establish a robust theoretical framework. This review spanned topics like human-centric and integrative lighting, the nuances of lighting asset management, and current lighting standards pertinent to libraries. Main results see section II-B).

#### 2) ON-SITE OBSERVATIONS

We conducted detailed observations at 20 selected Swedish public libraries, scrutinizing both conspicuous lighting challenges and the more insidious, often overlooked ones. At times, by positioning ourselves as library users, we indulged in participant observations, offering a firsthand experience of the lighting dynamics in these settings. Selection of the libraries are introduced in section II-C); main results refer to section III.

#### 3) STAKEHOLDER ENGAGEMENTS

Valuable insights were gleaned from interviews with a diverse cohort of stakeholders. This spectrum included lighting designers, academic researchers, energy managers, library managers, staff, and regular library users (referring to "users" in this study). Additionally, we executed a targeted questionnaire survey for the library managers, staff, and users, amassing over 100 responses during the summer of 2023. The interview design and execution refer to section II-D); main qualitative results refer to section IV.

Through these prongs, our methodology not only established the theoretical and practical foundations of the study but also facilitated an incisive gap analysis, pinpointing disparities between ideal lighting asset management practices and the existing ground realities.

### B. STATE-OF-THE-ART REVIEW

The evolution of lighting in public libraries has traversed centuries, progressing from natural daylight to open-flamed lighting, gaslight technology, and electric lighting. Today, the discourse surrounding library lighting remains vibrant and relevant [11]. Lighting, once perceived primarily from the function and a customer satisfaction perspective, has metamorphosed into an indispensable facet influencing the health and well-being of users [10], [12]. Well-designed and thoughtfully positioned lighting fosters user comfort and satisfaction, while inadequacies in lighting can curtail library visits and negatively impact both staff and users. The modern library's transformation into dynamic learning hubs amplifies the significance of appropriate lighting, emphasizing its critical role in enhancing human health and well-being.

The advent of LED technology ushered in an era of energy-efficient lighting solutions within libraries. Nevertheless, a crucial aspect often overlooked is the ongoing evaluation of lighting conditions, as most studies predominantly focus on the initial lighting design phase [13], [14]. Recent research trends have explored daylight integration in library building design and energy-efficient transitions to LED lighting, emphasizing strategies such as parallel, perpendicular, indirect, or hybrid lighting schemes in the public libraries [15], [16]. These approaches prioritize the infusion of ample natural light through strategic window placements and light redirection techniques, enhancing the overall library experience [17], [18]. Additionally, efficiency-focused measures include the transition to LED lighting, capitalizing on energy savings, prolonged lifespan, and customizable lighting options, as well as the integration of motion sensors to optimize energy usage [19].

Existing standards and guidelines pertaining to human-centric and integrative lighting asset management in public libraries encompass a spectrum of aspects, encompassing light and lighting standards [20], [21], technical memoranda [21], BIM (Building Information Modelling) properties for lighting [22], and dedicated asset management considerations. However, it is crucial to acknowledge certain limitations and challenges within this framework:

#### 1) RELIABILITY STANDARDS AND GUIDELINES
While standards for the reliability of traditional lighting sources in libraries exist, they often evaluate only 50% of the lamp's volume [23], leading to potential disparities in performance. Furthermore, guidelines for LED reliability are typically anchored to the "L70" benchmark [24], which can be restrictive and not fully reflective of real-world scenarios. This can result in misinterpretations and misapplications of these standards by users and lighting asset managers.

#### 2) LIMITED PARAMETERS
The parameters identified for public libraries within these standards and guidelines are relatively limited in scope, predominantly focusing on illumination levels and related metrics. This limited scope may not encompass the multi-faceted lighting needs and preferences of library users.

This study has identified examples of lighting performance parameters in public libraries as shown in FIGURE 1, which refer to: illuminance, glare, color rendering, color temperature, flicker, chromaticity, energy efficiency, life span, and other factors. These parameters can provide a comprehensive view of a lighting system's performance and how well it meets the needs and preferences of users.

#### 3) APPLICATION CONTEXTS
The application contexts specified in these standards are primarily centred around bookshelves, reading areas, counters, and general lighting. However, these predefined contexts may not authentically reflect the diverse and dynamic ways in which library patrons interact with their environment. Users may engage with books at various heights, sit in different areas, or utilize various pieces of furniture, each requiring unique lighting considerations. These standards often specify measurements at fixed heights or distances from walls and ceilings, which may not align with the nuanced, real-world usage patterns of library visitors.

Moreover, studies examining the post-design, utilization phase of library lighting remain sparse. Limited investigations delve into aspects such as the assessment of natural lighting and visual comfort within library spaces or comparative analyses of daylighting versus artificial lighting in library buildings [25], [26]. It is imperative to note that while a significant number of studies have concentrated on libraries located at latitudes of 5 to 35 degrees, there exists a gap in research for libraries positioned at higher latitudes, specifically over 55 degrees from the equator. This gap is of considerable importance as it presents an opportunity to investigate the distinct lighting challenges and considerations that are inherent to these northern latitudes. Furthermore, only a handful of studies scrutinize the energy usage patterns of libraries [27]. A notable gap exists concerning the effectiveness of lighting from human-centric and integrative perspectives once library buildings are in active use, regardless of their age. Consequently, there is a need for a paradigm shift towards human-centric and integrative lighting asset management in public libraries, placing renewed emphasis on the continual monitoring and review of both visual and non-visual lighting effects.

The deficiency in lighting asset management extends beyond just library buildings; it is a prevalent concern across a diverse spectrum of structures, often operating reactively — addressing issues only when they arise visibly, leading to prolonged maintenance time, higher costs, inefficiencies resource and energy. Industry trailblazers have introduced concepts like lighting-as-a-service (LaaS) [28] and initiatives like Lighting4People [10]. However, the practical applicability of these innovative approaches has not yet been thoroughly examined in the specific context of libraries.

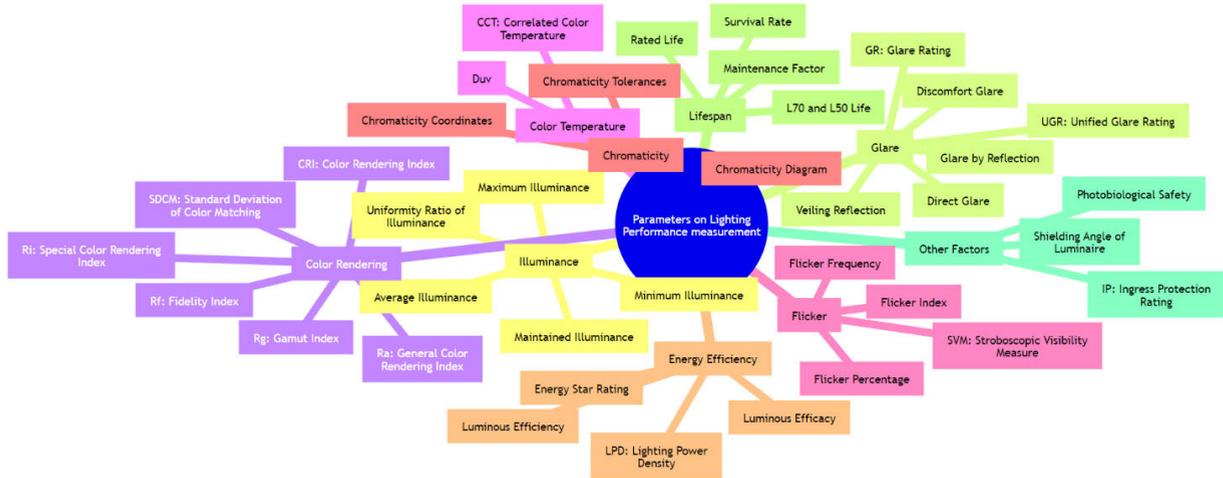

**FIGURE 1.** Lighting performance parameters.

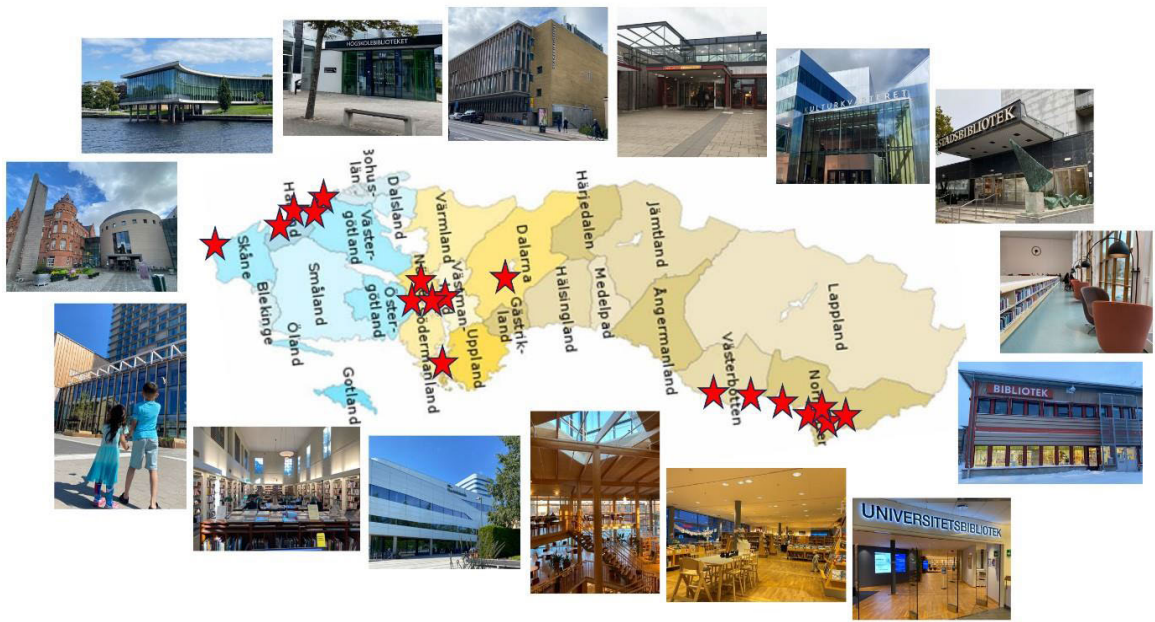

**FIGURE 2.** Libraries in a Swedish field study.

In summary, state-of-the-art review from this study reveals a crucial gap exists in ongoing lighting evaluation during active library use, especially from human-centric and integrative perspectives. It also reveals the gap in existing standards, though comprehensive, exhibit limitations in reliability assessments and parameters, failing to fully capture the diverse needs of library users and their dynamic interactions with the environment.

### C. SELECTION OF THE SWEDISH PUBLIC LIBRARIES FOR ON-SITE OBSERVATIONS

A study was undertaken across 20 public libraries (incl. 5 University libraries) spread over diverse Swedish cities, from the coastal city of Luleå in the north to the cultural epicenter of Malmö in the south. This study spanned from October 2022 to July 2023, meticulously covering libraries in Luleå, Piteå, Skellefteå, Umeå, Stockholm, Falun, Eskilstuna, Örebro, Västerås, Gothenburg, Borås, Borlänge, Halmstad, and Malmö (see FIGURE 2; photos are not available from all libraries visited). The covered latitude range is from 55.6050° N (Malmö) to 65.5848° N (Luleå).

### D. INTERVIEW DESIGN AND EXECUTION WITH STAKEHOLDERS (MANAGERS, STAFF, USERS)

In our extensive field study encompassing 20 public libraries, our research methodology extended beyond on-site observations to include interviews with engaged stakeholders, comprising library managers, staff, and users keen to share their valuable insights. Our approach initiated with an introduction to the focal point of our study, the intricate realm of lighting systems, followed by a clear exposition of our interest in human-centric and integrative lighting principles and their benefits. We encouraged participants to candidly express their user experiences and perspectives.

Moreover, we thoughtfully designed three distinct questionnaires tailored to different stakeholders: library managers, staff, and users. The questionnaire directed towards library managers sought their feedback on the implementation and maintenance of human-centric and integrative lighting in libraries. Given their pivotal role in the absence of dedicated lighting asset managers, their insights were instrumental in understanding the challenges, benefits, and best practices essential for crafting a comfortable and productive environment for both staff and users. The questionnaire aimed at library staff facilitated the collection of feedback on their experiences with human-centric and integrative lighting, offering a crucial avenue for enhancing lighting conditions and ensuring comfort and productivity. Similarly, the questionnaire for users solicited feedback to gain insights into their experiences with human-centric and integrative lighting, contributing to our ongoing efforts to refine lighting conditions for the betterment of all users.

However, it is crucial to acknowledge the limitations of this paper, as it solely presents qualitative findings obtained from interviews and questionnaires. Moreover, the limited number of responses precludes the drawing of statistically significant, long-term conclusions.

## III. QUALITATIVE INSIGHTS FROM ON-SITE INVESTIGATIONS

Section III delves into the empirical findings from our field study on lighting in Swedish public libraries. We spotlight best practices in human-centric lighting, highlight visible and invisible problems, and we assess the present efficacy of lighting asset management and its challenges.

### A. OBSERVATIONS ON HUMAN-CENTRIC AND INTEGRATIVE LIGHTING

This section presents the findings from observations of our study, identifying best practices and categorizing the unearthed issues, which includes both overt problems and covert challenges.

#### 1) BEST PRACTICES IDENTIFIED

Throughout our meticulous observations, several noteworthy best practices in the realm of human-centric and integrative lighting have come to the fore, spanning various aspects:

*a: LED APPLICATIONS*

In newer libraries, the effective integration of LED lighting solutions stands out prominently. This implementation augments energy efficiency and thereby contributes to reducing the carbon footprint.

*b: FENESTRATION (FULL GLAZED FAÇADE/ATRIUM/CLERESTORY)*

Incorporating glass walls, ceilings, and windows within library structures provides a notable advantage. These architectural elements enable the effective utilization of natural daylight, consequently optimizing energy consumption. Notably, some libraries have judiciously utilized curtains to mitigate excessive sunlight when required.

*c: SUPPLEMENTARY LIGHTING WITH USER CONTROL*

Some library spaces are equipped with tables and sofas that incorporate user self-adjustable lighting. This feature empowers users to tailor lighting conditions to their preferences, ensuring a personalized and comfortable experience.

*d: BOOKSHELF DESIGN*

One approach to bookshelf design considers the strategic placement of lighting systems. For instance, shelves with sloping surfaces enable the top-level lighting to illuminate lower levels, ensuring an efficient use of light. Additionally, lighting installations in bookshelves, on each shelf level, further enhancing the reading experience.

*e: DYNAMIC FURNITURE ARRANGEMENT*

An adaptable approach to furniture placement leverages the changing position of furniture with respect to daylight. As we visited libraries during different seasons, it became evident that furniture rearrangement was employed to maximize natural daylight utilization, offering an energy-saving solution.

These best practices in TABLE 1 underscore the synergy between lighting design and human health and well-being, ultimately contributing to a more sustainable library environment.

#### 2) AREAS NEEDING ENHANCEMENT: VISIBLE AND INVISIBLE PROBLEMS

This section aims to investigate the problems associated with current lighting performance that are visible or invisible to the naked eye. These problems, which range from catastrophic failure to inefficient communication, have significant implications on the overall lighting quality and user experience. By addressing these concerns, better lighting solutions can be implemented for improved comfort and efficiency.

As shown in TABLE 2, those overt problems include: *Catastrophic Failure*: Sudden and complete breakdowns in the lighting system can lead to unfunctional light sources and compromised illumination; *Parametric Degradation of light sources*: The gradual deterioration of lighting performance over time, leading to reduced brightness and colour accuracy; *Flicker*: Inconsistent or fluctuating light output that can be visually disturbing and cause discomfort for occupants; *Inappropriate Operation by Manager or User*: Improper use or handling of lighting equipment, leading to suboptimal performance or damage; *Lighting Design Issues*: Lighting solutions that do not adequately cater to the specific needs and requirements of a space; *Outdated Technology*: The continued use of obsolete lighting technologies that are less efficient and reliable compared to modern alternatives; *Inefficient Communication with Building Designer*: Lack of effective collaboration between lighting and building design professionals, resulting in subpar integration and coordination; *Dirt and Contamination*: Accumulation of dirt, dust, and other contaminants on lighting fixtures and surfaces, leading to reduced light output and compromised aesthetics; *Excessive and unnecessary lighting*: such lighting in public libraries can lead to energy inefficiency and user discomfort.

**TABLE 1.** Best practices identified: some examples.

| BEST PRACTICES | ILLUSTRATION PHOTOS TAKEN FROM OBSERVED PUBLIC LIBRARIES IN SWEDEN |
|---|---|
| LED APPLICATIONS | |
| FENESTRATION: FULL GLAZED FACADE | |
| FENESTRATION: ATRIUM/CLERESTORY | |
| SUPPLEMENTARY LIGHTING WITH USER CONTROL | |
| BOOKSHELF DESIGN | |

**TABLE 2.** Overt problems: some examples.

| PROBLEMS | ILLUSTRATION PHOTOS TAKEN FROM OBSERVED PUBLIC LIBRARIES IN SWEDEN |
|---|---|
| CATASTROPHIC FAILURE | 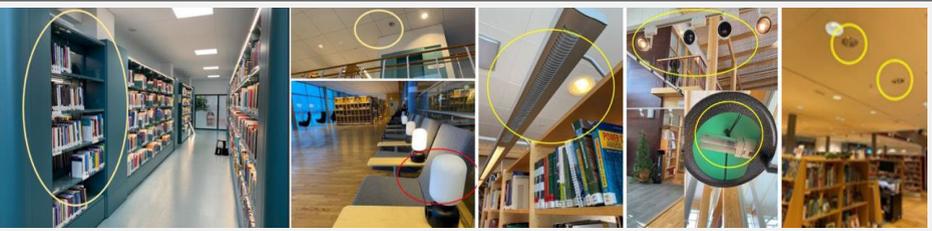 |
| PARAMETRIC DEGRADATION OF LIGHT SOURCES | 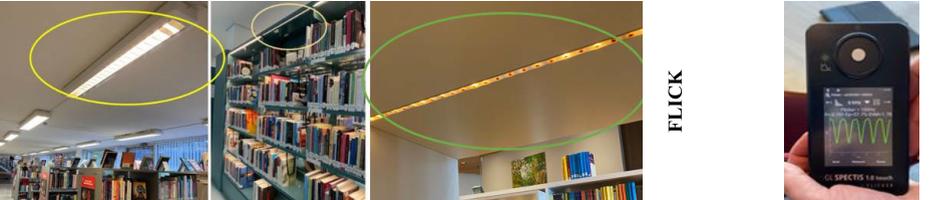 |
| INAPPROPRIATE OPERATION BY MANAGER OR USER | 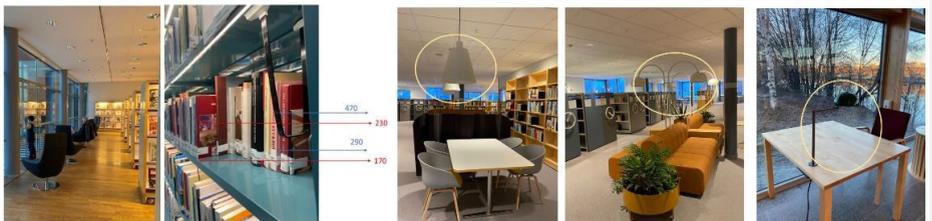 |
| LIGHTING APPLICATION | 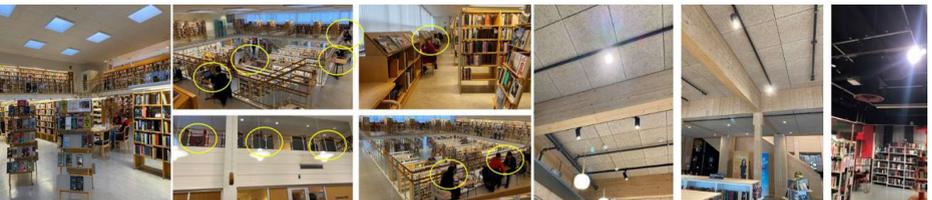 |
| OUTDATED TECHNOLOGY | 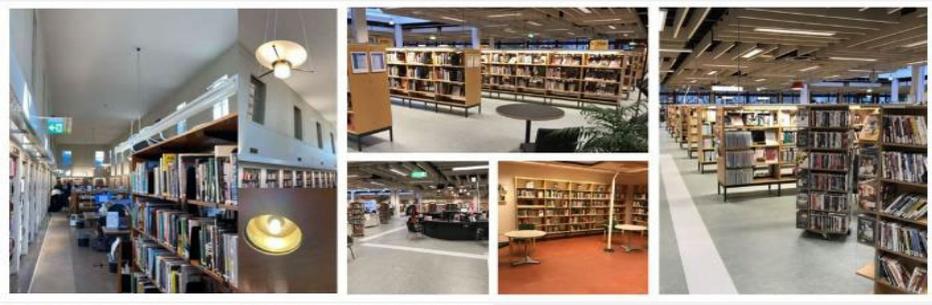 |
| INEFFICIENT COMMUNICATION WITH BUILDING DESIGNER / EXCESSIVE OR UNNECESSARY LIGHTING | 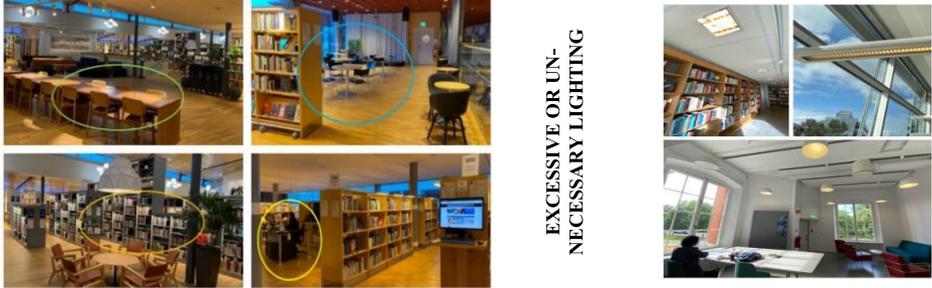 |

Addressing these prevalent lighting problems is crucial for creating human-centric and integrative lighting. By identifying and tackling these issues, stakeholders can optimize lighting systems to enhance user experiences and overall indoor environment (building) performance.

*a: CATASTROPHIC FAILURE*
It refers to a significant and sudden breakdown of a system, is one of the primary causes of unfunctional light sources. This type of failure can affect all kinds of light sources, including LED and non-LED types, as well as libraries of various ages, from older establishments to newer facilities. It is crucial to examine and address the issue of catastrophic failure to prevent unfunctional lighting fixtures in libraries. By understanding the underlying causes, appropriate preventive measures can be implemented to ensure a well-lit and conducive environment for learning and research in all libraries, regardless of their type or age.

#### b: TYPICAL PARAMETRIC FAILURE—LIGHT SOURCES

It refers to the gradual decline in the performance of a component or system over time, which can ultimately lead to its malfunction. In the context of lighting, parametric failure can result in unfunctional bulbs.

Degradation of light sources is a key manifestation of parametric failure in lighting systems. This issue is prevalent in many types, including LED and non-LED varieties, and affects libraries of all ages, from older establishments to newer facilities. The degradation is characterized by a reduction in brightness, colour accuracy, and overall lighting performance.

Understanding and addressing the impact of parametric failure on unfunctional light fixtures is essential for maintaining optimal lighting conditions in libraries. By identifying the causes and signs of degradation, appropriate preventive measures and maintenance strategies can be implemented to ensure a well-lit and conducive environment for learning and research in libraries of all types and ages.

#### c: FLICKER

It refers to a significant issue in lighting systems, causing inconsistent and fluctuating light output. This inconsistency not only detracts from the overall lighting quality but also has a negative impact on user experience. Flicker is not only annoying but also potentially harmful to occupants. It can cause discomfort, visual disturbances, headaches, and even trigger seizures in individuals with photosensitive epilepsy.

During the investigation, it was discovered that some instances of visible flickering were particularly obvious and had persisted for at least six months, or even longer. The prolonged presence of flicker can exacerbate its detrimental effects on users, further emphasizing the need for timely identification and resolution of this issue.

Invisible flicker refers to rapid fluctuations in light output that are not perceptible by the human eye. Despite being undetectable, this flicker can cause discomfort, eye strain, and even headaches for some individuals. Identifying and addressing invisible flicker in lighting systems can significantly improve user comfort. During the investigation, some invisible flicker also can be detected.

Understanding and addressing the issue of flicker is crucial for maintaining optimal lighting performance and user experience. By promptly identifying and rectifying instances of flicker, stakeholders can ensure a comfortable and safe environment for occupants, minimizing the negative effects of this common lighting failure.

#### d: INAPPROPRIATE OPERATION BY LIGHTING OPERATOR

When library operators mishandle or misuse lighting equipment, it can result in reduced performance or even damage to the systems. This can negatively impact the overall quality of lighting and hinder the library's functionality.

Operators may adjust lighting settings to save energy. However, without proper knowledge and training, these adjustments could lead to suboptimal lighting conditions that impair the user experience and may not actually result in significant energy savings.

Operators may lack the necessary knowledge, or the inflexible lighting systems are not suited for a library with its changing layout, which can lead to improper furniture/book positioning that adversely affects the quality of illumination.

Particularly in higher latitudes, architectural approaches that adapt to the unique challenges of the winter season are essential due to the extreme variability of sunlight, ranging from the midnight sun in summer to prolonged darkness in winter. An integrated approach to lighting should be comprehensive, encompassing quality views, shading systems, sufficient illumination, glare analysis, and multidirectional aperture configuration.

Addressing the issue of improper lighting equipment handling in public libraries is crucial for maintaining an optimal lighting environment that promotes a positive user experience. By providing appropriate training and education to library staff, who are often responsible for various operational tasks, stakeholders can ensure that lighting systems are used effectively and efficiently, ultimately enhancing the library experience for all users.

#### e: INAPPROPRIATE OPERATION BY USER

The issue of inappropriate operation of library lighting systems by users can lead to dark lighting condition. Factors such as personal preferences and user-unfriendly lighting systems contribute to this problem. Instances where users are either unaware of their responsibility to turn on the lights or unable to do so due to lack of instructions should also be noticed.

#### f: LIGHTING APPLICATION ISSUES

Factors contributing to lighting application issues include insufficient consideration of daylight, user needs, and proper lighting properties. The report highlights specific issues, such as the improper use of spot lighting, which can result in glare and negatively impact both library staff and users. Causes of poor lighting application in Public Libraries include:

1) Lack of User Consideration: Poor lighting design may not prioritize the needs of library users, resulting in a lighting approach which is not human-centric. This can lead to an environment that is not conducive to reading, studying, or engaging in other library activities.
2) Improper Lighting Application: In some cases, poor lighting application may involve the inappropriate use of specific lighting properties, such as improperly implemented spot lighting. This can cause annoying glares for both library staff and users, negatively affecting their experience within the library.

Addressing the issue of poor lighting application in public libraries is essential for creating a human-centric environment that promotes a positive user experience. By considering factors such as daylight, user needs, and appropriate lighting properties, libraries can optimize their lighting application and provide a comfortable, well-lit space for all users.

### g: OUTDATED TECHNOLOGY

The impact of persistently using outdated lighting technologies that are less efficient and reliable compared to their modern counterparts would have negative impact on energy efficiency and human health. Specific issues examined include the utilization of energy inefficient light sources, which contribute to increased heat during the summer, and the challenges of sourcing replacement light sources that are no longer available in the market. Issues with obsolete lighting technologies include:

1) Energy Inefficient Light Source: Continuing to use outdated, energy inefficient light source can lead to excessive heat generation, particularly during the summer months. This not only increases energy consumption and costs but also creates an uncomfortable environment for users.
2) Unavailability of Replacement Light Source: As older lighting technologies become obsolete, sourcing replacement light source becomes increasingly difficult. When these light sources are no longer available in the market [29], maintaining and repairing the lighting system can become a significant challenge.

Addressing the issue of continued use of obsolete lighting technologies is essential for optimizing energy efficiency, reliability, and user experience in modern environments. By transitioning to more efficient and readily available lighting technologies, stakeholders can ensure improved performance and sustainability while providing a comfortable, well-lit space for all users.

### h: INEFFICIENT COMMUNICATION BETWEEN LIGHTING AND BUILDING DESIGN PROFESSIONALS

As observed in the study, the consequences of inefficient communication between lighting and (new or existing) building design professionals may result in suboptimal integration and coordination. The lack of effective collaboration can lead to a range of issues, including incorrect furniture positioning, improper lighting design for users, unsuitable lighting source election, and an absence of continuous improvement when addressing new concerns. Effects of inefficient communication include:

1) Incorrect Furniture Positioning: Poor collaboration between lighting and building design professionals can result in improperly positioned furniture, potentially compromising the overall functionality and user experience within the space.
2) Improper Lighting Design for Users: Inefficient communication may lead to lighting designs that do not adequately cater to users' needs, negatively impacting their comfort and ability to effectively use the space.
3) Unsuitable Lighting Fixture and/or Lighting Source Selection: A lack of effective collaboration can also result in the selection of inappropriate light sources and/or fixtures, leading to improper lighting conditions and a higher likelihood of light sources failing prematurely.
4) Absence of Continuous Improvement: When communication between lighting and building design professionals is inefficient, it becomes difficult to promptly identify and address new issues that arise, hindering the process of continuous improvement and adaptation. Adjusting lighting in libraries to suit evolving space needs is challenging due to high costs and often, lack of expertise. While adaptable lighting could help, it requires careful management and can be limited by building constraints in historic venues. Thus, lighting issues are not just about design foresight but also about managing practical constraints and resources.

Fostering efficient communication between lighting and building design professionals is crucial for creating well-integrated, functional spaces that cater to user needs. By improving collaboration, stakeholders can ensure that lighting systems are properly designed and maintained, ultimately enhancing the overall user experience within the built environment.

### i: EXCESSIVE AND UNNECESSARY LIGHTING

As observed in the study, because of the extreme variability of the solar position in higher latitude locations across Sweden, excessive and unnecessary lighting in public libraries can lead to energy inefficiency and user discomfort. The analysis from the investigation covers various factors contributing to overly bright lighting conditions, including lights left on when not in use, excessive daylight during specific times, and deviations from standard recommendations. Causes of excessive and unnecessary lighting in Public Libraries include:

1) Lighting Left on When Not in Use: Failing to turn off lights when they are not needed in public libraries contributes to energy inefficiency and excessively bright environments. Ensuring that lights are switched off when not in use can help reduce energy consumption and provide a more comfortable atmosphere for users. However, it should be noted that completely turning off lights in unoccupied spaces can lead to an unwelcoming atmosphere that may deter individuals from entering. The study suggests that dimming lights to a level of 10-25% when spaces are not in use strikes a better balance, creating a more inviting environment [30].
2) Excessive Daylight During Daytime or Summer Months: During the day or in summer months, daylight may be more than sufficient for indoor illumination in libraries. Over-reliance on artificial lighting in these situations can lead to overly bright environments and energy inefficiency.

3) Over-lighting Due to Snow Reflection in Winter: In winter, snow outside library windows can reflect sunlight and cause over-lighting indoors. This excess illumination can lead to discomfort and unnecessary energy consumption if not managed properly. Reducing discomfort from glare caused by sunlight reflecting off snow can be mitigated if indoor lighting is brighter, aiding in eye adjustment. However, when this is not feasible, the use of curtains to dim the incoming light becomes necessary, and supplementary artificial lighting must still be employed to maintain visibility.

4) Exceeding Lighting Design Recommendations: Over-illumination may occur when indoor library environments are lit beyond the recommended levels, resulting in unnecessary brightness and energy consumption. Adhering to lighting design recommendations can help maintain optimal lighting conditions that are both comfortable and energy efficient.

Addressing the issues of excessive and unnecessary lighting is crucial for creating comfortable and energy-efficient public library environments. By being mindful of lighting usage, taking advantage of natural daylight, and adhering to standard recommendations, library operators can create well-lit spaces that are both energy-efficient and user-friendly.

### B. EFFECTIVENESS AND CHALLENGES OF LIGHTING ASSET MANAGEMENT

Our on-site observations and active stakeholder engagements offer profound insights into the effectiveness and challenges entailed in human-centric and integrative lighting asset management within studied public libraries. To comprehensively assess these dynamics, we conducted a gap analysis from the vantage point of asset management, comprising four critical facets: Plan & Implementation, Study & Reflection, Act & Optimization, and Digitalization & Visualization.

An overview of the findings from our investigation across the studied public libraries is succinctly presented in Table 3.1. This section underscores whether the current lighting asset management practices have indeed kept pace with technological advancements, particularly in the context of LED lighting solutions.

#### 1) PLAN & IMPLEMENTATION

The analysis on Plan & implementation highlights areas of concern, including the lack of clear policies, a focus on reactive maintenance, and insufficient consideration of visual and non-visual effects. Developing clear policies, prioritizing preventive maintenance, and considering both visual and non-visual effects can help libraries improve their lighting solutions and enhance the overall user experience. Current challenges in lighting asset management from Plan & Implementation include:

*a: LACK OF CLEAR POLICY OR STRATEGY*
Public libraries often lack a well-defined policy or strategy for managing lighting assets, leading to inefficiencies and suboptimal lighting conditions.

*b: FOCUS ON CATASTROPHIC FAILURES*
Existing strategies tend to prioritize addressing catastrophic failures rather than proactively maintaining and optimizing lighting systems.

*c: REACTIVE MAINTENANCE STRATEGY*
A "fix it only if it breaks" approach to maintenance results in longer downtimes and increased costs, as opposed to adopting preventive maintenance practices.

*d: SLOW RESPONSE TIME*
Libraries may take an extended period (over 12 months) to address lighting failures, leading to user dissatisfaction, and compromised lighting conditions.

*e: LIMITED PREVENTIVE MAINTENANCE*
While preventive maintenance might be considered occasionally, it is not consistently applied or prioritized in lighting asset management.

*f: ABSENCE OF MONITORING AND PREDICTION*
Public libraries often do not monitor or predict visual or non-visual effects, leading to a lack of data-driven decision-making in lighting management.

*g: NEGLECT OF ENERGY CONSUMPTION*
Energy consumption is rarely considered when managing lighting assets, which can result in higher operating costs and negative environmental impacts.

*h: SEPARATE MANAGEMENT OF BUILDING AND BOOK LIGHTING*
The lighting of library buildings and lighting for books are often managed separately, leading to inconsistencies and unclear responsibilities.

*i: LACK OF RISK MANAGEMENT PLAN*
Public libraries typically do not have a risk management plan in place for their lighting assets, leaving them unprepared for potential issues and disruptions.

#### 2) STUDY & REFLECTION
The analysis on Study & Reflection highlights areas of concern, including focusing on aspects such as lighting function, energy efficiency, adherence to standards and guidelines, and human-centric objectives. Identifying areas of concern can help improve lighting solutions and create more comfortable library environments. Key findings in some investigated library, include:

*a: LIGHTING FUNCTION*

1) Some catastrophically failed lights remain unreplaced after 12 months.
2) New catastrophic failures occurred continuously during the 3-week study period.
3) Some lighting sources require replacement due to noticeable colour shifts or flickering.
4) Some lights are improperly installed or used.
5) Energy Efficiency
6) The designed number and type of lights may be unsuitable.
7) Over-illumination is observed in some areas.
8) Natural light is not sufficiently utilized.
9) Lighting fixture selection needs re-evaluation, even for libraries with LED lighting.
10) Failed light sources may still consume energy.
11) Aging lights (without obvious failure) may consume more power than designed.
12) Older libraries have more non-LED lighting installed.

*b: STANDARDS AND GUIDELINES*

1) Reliability definitions for light sources have limitations, evaluating only 50% of their volume.
2) The 'L70' benchmark is commonly used to define the reliability of LED lighting, indicating the point at which light output drops to 70% of its original luminosity. However, users have the flexibility to select a higher 'L' value as a benchmark for increased longevity, according to their preferences.
3) Lighting standards specify a general measurement height of 0.75m. However, this standard may not adequately represent other essential requirements, potentially leading to underestimations in various lighting contexts.
4) Maintenance factors are not adequately considered.

*c: INTEGRATIVE AND HUMAN-CENTRIC OBJECTIVES*

1) Light intensities are unbalanced throughout the library, resulting in stark illumination contrasts in connected reading areas, which can lead to visual discomfort.
2) Limited knowledge on selecting and using appropriate lighting sources and fixtures by library staff and users.
3) Limited consideration of user context, ex. varying ages, heights, activities, sunlight intensity, etc.
4) Continuous monitoring, prediction, and optimization are lacking.
5) 70% of measured points (approximately 30 measurement points in total) demonstrated insufficient light intensity levels compared to human well-being standards.

3) ACT & OPTIMIZATION

The analysis on Act & Optimization highlights areas of concern, focusing on aspects related to optimization, research data, warranty and field data analysis, and integration with other asset management strategies. Key reflections through investigation include:

*a: LACK OF OPTIMIZATION*

There is no clear optimization strategy for lighting asset management in studied public libraries.

*b: ABSENCE OF WARRANTY AND FIELD DATA ANALYSIS*

Studied libraries do not analyze warranty or field data to inform their lighting asset management decisions.

*c: NO FRACAS IMPLEMENTATION*

Public libraries lack a failure reporting, analysis, and corrective action system (FRACAS) for lighting asset management.

*d: INCONSISTENT MAINTENANCE STRATEGIES*

Lighting asset management is not consistently integrated with other asset management strategies, such as ventilation.

4) DIGITALIZATION & VISUALIZATION

Introducing digitalization and visualization solutions to lighting asset management in public libraries can lead to improved lighting systems and overall user experience. By adopting these technologies, libraries can optimize their lighting assets more effectively and create more comfortable and efficient environments. However, current investigations reveal that most studied public libraries do not utilize digitalization or visualization tools for managing their lighting assets.

## IV. QUALITATIVE INSIGHTS FROM STAKEHOLDER INTERVIEWS AND QUESTIONNAIRES

As introduced in Section II, in our extensive field study of 20 public libraries, we conducted interviews and surveys with stakeholders, including library managers, staff, and users, to gather insights into human-centric and integrative lighting. We introduced the study's focus on lighting systems and encouraged participants to share their experiences and perspectives. Three tailored questionnaires were designed for library managers, staff, and users, gathering feedback on lighting implementation and user experiences.

In total, our study gathers over 150 feedback, among which, there are 3 from managers, 23 from staff, and 125 from users. This paper exclusively presents qualitative results from these interviews and questionnaires, offering valuable insights into the perspectives of library stakeholders.

### A. MANAGERIAL INSIGHTS AND PRIORITIES

Based on the few interviews and questionnaires with library managers, some key insights and priorities regarding human-centric and integrative lighting asset management emerge:

1) DISSATISFACTION WITH CURRENT MANAGEMENT

All three library managers express wishes to improve the current state of lighting asset management in their libraries.

#### 2) LACK OF DESIGN KNOWLEDGE
Library managers are generally unfamiliar with the design specifications and requirements for lighting, including illuminance thresholds, flicker, and colour temperature standards.

#### 3) ABSENCE OF LIGHTING STRATEGY
All three library managers missed developing a dedicated lighting asset management strategy for their libraries, leading to a lack of systematic review or optimization efforts.

#### 4) POOR DOCUMENTATION AND COMMUNICATION
Processes related to lighting management and maintenance lack proper documentation and communication with library staff.

#### 5) COMMON LIGHTING ISSUES
Staff members have reported common lighting issues such as excessive brightness, and a lack of adjustability.

#### 6) ROOT CAUSE ANALYSIS
No library manager has conducted a comprehensive root cause analysis to address undesirable lighting conditions systematically.

#### 7) LACK OF LIGHTING ASSESSMENT
None of the managers have conducted a formal lighting assessment within their libraries.

#### 8) CHALLENGES IN MANAGEMENT
Challenges faced in lighting asset management include budget constraints, technical difficulties, and the need for maintenance and upgrades.

#### 9) UNFAMILIARITY WITH HUMAN-CENTRIC LIGHTING
The library managers show a general unfamiliarity with the concept of human-centric and integrative lighting.

#### 10) UNAWARENESS OF ISO 55000 STANDARDS
None of the managers are acquainted with ISO 55000 standards, which provide guidelines for asset management.

#### 11) DESIRE FOR GUIDANCE
All managers express a desire for more guidance on managing lighting to support the health and well-being of both staff and users.

#### 12) INFORMAL DECISION-MAKING
Decision-making related to lighting asset management is primarily based on individual experiences and intuition.

#### 13) LACK OF FORMAL IMPROVEMENT INITIATIVES
There are no formal initiatives in place for continuous improvement in lighting asset management within the libraries.

In conclusion, the findings in TABLE 3 would be beneficial to enhanced lighting asset management practices in public libraries, with a focus on knowledge acquisition, strategy development, documentation, and formalized improvement efforts. Additionally, there is a clear desire from the library managers for guidance in human-centric lighting asset management to benefit both staff and users.

### B. FEEDBACK AND PERSPECTIVES FROM LIBRARY STAFF
Based on interviews and questionnaires with library staff, several key feedback, and perspectives regarding human-centric and integrative lighting asset management in public libraries can be summarized:

#### 1) GENERAL LIGHTING EXPERIENCE
Most library staff perceive the current lighting conditions in the library as average, falling neither in the category of "very good" nor "very bad".

Many staff members have experienced eye strain, discomfort, or fatigue due to various lighting conditions within the library, including counters, inquiry tables with multimedia, bookshelves, and office rooms. These discomforts are often associated with factors such as brightness, darkness, glare, flicker, excessively cold colour temperature, or the inability to adjust lighting.

#### 2) LIGHTING CONCERNS AND PREFERENCES
A majority of staff members prefer lighting options that are adjustable to accommodate personal preferences and task-specific needs.

Half of the staff have reported lighting issues to library managers, with most issues taking more than four weeks to be addressed. Commonly reported issues include broken light sources, inadequate lighting, glare, and the inability to adjust lighting.

Only one-third of staff members express a willingness to communicate with library managers about their lighting environment experiences, despite half of them having received lighting complaints from users.

#### 3) STAFF EXPECTATIONS AND SUGGESTIONS
About half of the staff are interested in receiving guidance from the library regarding lighting choices that support human health and well-being. Specific suggestions include directional lighting, flexible lighting solutions, movable spotlights, dimmers, and options for creating cozy lighting environments in spaces like the fairytale room.

#### 4) HUMAN-CENTRIC AND INTEGRATIVE LIGHTING CONCERNS
Most staff members are not familiar with the concept of human-centric and integrative lighting.

Nearly all staff members acknowledged the significance of human-centric and integrative lighting; upon receiving relevant information, they considered it to be beneficial and quite important.

**TABLE 3.** Overview from investigations in studied public libraries.

| ITEMS | SUBITEMS | REFLECTIONS |
|---|---|---|
| PLAN & IMPLEMENTATION | / | • Lack of Clear Policy or Strategy: Public libraries often lack a well-defined policy or strategy for managing lighting assets, leading to inefficiencies and suboptimal lighting conditions.<br>• Focus on Catastrophic Failures: Existing strategies tend to prioritize addressing catastrophic failures rather than proactively maintaining and optimizing lighting systems.<br>• Reactive Maintenance Strategy: A "fix it only if it breaks" approach to maintenance results in longer downtimes and increased costs, as opposed to adopting preventive maintenance practices.<br>• Slow Response Time: Libraries may take an extended period (over 12 months) to address lighting failures, leading to user dissatisfaction and compromised lighting conditions.<br>• Limited Preventive Maintenance: While preventive maintenance might be considered occasionally, it is not consistently applied or prioritized in lighting asset management.<br>• Absence of Monitoring and Prediction: Public libraries often do not monitor or predict visual or non-visual effects, leading to a lack of data-driven decision-making in lighting management.<br>• Neglect of Energy Consumption: Energy consumption is rarely considered when managing lighting assets, which can result in higher operating costs and negative environmental impacts.<br>• Separate Management of Building and Book Lighting: The lighting of library buildings and lighting for books are often managed separately, leading to inconsistencies and unclear responsibilities.<br>• Lack of Risk Management Plan: Public libraries typically do not have a risk management plan in place for their lighting assets, leaving them unprepared for potential issues and disruptions. |
| STUDY & REFLECTION | Lighting Function | • Some catastrophically failed lights remain unreplaced after 12 months.<br>• New catastrophic failures occurred continuously during the 3-week study period.<br>• Some lighting sources require replacement due to noticeable colour shifts or flickering.<br>• Some lights are improperly installed or used. |
| | Energy Efficiency | • The designed number and type of lights may be unsuitable.<br>• Over-illumination is observed in some areas.<br>• Natural light is not sufficiently utilized.<br>• Lighting sources and fixtures selection needs re-evaluation, even for libraries with LED lighting.<br>• Failed light sources may still consume energy.<br>• Aging lights (without obvious failure) may consume more power than designed.<br>• Older libraries have more non-LED lighting installed. |
| | Standards & Guidelines | • Reliability definitions for light sources have limitations, evaluating only 50% of their volume.<br>• The 'L70' benchmark is commonly used to define the reliability of LED lighting, indicating the point at which light output drops to 70% of its original luminosity. However, users have the flexibility to select a higher 'L' value as a benchmark for increased longevity, according to their preferences.<br>• Lighting standards specify a general measurement height of 0.75m. However, this standard may not adequately represent other essential requirements, potentially leading to underestimations in various lighting contexts.<br>• Maintenance factors are not adequately considered. . |
| | Integrative/ Human-centric objectives | • Light intensities are unbalanced throughout the library, resulting in stark illumination contrasts in connected reading areas, which can lead to visual discomfort.<br>• Limited knowledge on selecting and using appropriate lighting exists.<br>• User context (varying ages, heights, activities, sunlight intensity, etc.) is insufficiently considered.<br>• Continuous monitoring, prediction, and optimization are lacking.<br>• 70% of measured points (approximately 30 measurement points in total) demonstrated insufficient light intensity levels compared to human well-being standards. |
| ACT & OPTIMIZATION | / | • Lack of Optimization: There is no clear optimization strategy for lighting asset management in studied public libraries.<br>• Absence of Warranty and Field Data Analysis: Studied libraries do not analyze warranty or field data to inform their lighting asset management decisions.<br>• No FRACAS Implementation: Public libraries lack a failure reporting, analysis, and corrective action system (FRACAS) for lighting asset management.<br>• Inconsistent Maintenance Strategies: Lighting asset management is not consistently integrated with other asset management strategies, such as ventilation. |
| DIGITALIZATION & VISULATION | / | • Absence of Digitalization and Visualization Solutions for lighting asset management in studied libraries. |

Important aspects of human-centric and integrative lighting identified by staff include adjustable brightness levels, lighting that responds to natural daylight, and task-specific lighting options, particularly for special workshops.

In conclusion, library staff feedback highlights the importance of addressing lighting issues to create a more comfortable and productive environment for both staff and users. Their willingness to receive guidance on lighting choices that support human health and well-being underscores the potential for improvement in lighting asset management practices.

While most staff are not yet familiar with the terminology, their recognition of the importance of human-centric and integrative lighting suggests an opportunity to bridge this knowledge gap and implement lighting solutions that better meet the needs of library users.

### C. LIBRARY USERS' EXPERIENCES, PREFERENCES, AND SUGGESTIONS

Based on interviews and questionnaires with library regular users, the experiences, preferences, and suggestions regarding human-centric and integrative lighting asset management in public libraries can be summarized as follows:

### 1) GENERAL LIGHTING EXPERIENCE:

The majority of library users' express satisfaction with the current lighting conditions in the library, with a smaller portion considering it average and falling between ''very good'' and ''very bad.''

Two-thirds of users believe that the lighting conditions support their ability to read and concentrate effectively.

### 2) LIGHTING CONCERNS AND PREFERENCES:

Some users have experienced eye strain or discomfort due to specific lighting conditions in the library, especially during sunny weather (with/without curtains), overcast skies, and polar nights in the winter. These discomforts are primarily attributed to lighting being too dim or the inability to adjust it, while a smaller portion experience glare or flicker.

Half of the users prefer adjustable lighting options to better align with their personal preferences and tasks. Areas identified for improvement include newspaper/magazine reading areas and book borrowing/returning areas.

### 3) USER EXPECTATIONS AND SUGGESTIONS

Even when users are dissatisfied with the lighting conditions, they are not comfortable to communicate their lighting environment experiences with library staff or seek additional guidance from the library regarding lighting choices that support their health.

Most users are not familiar with the concept of human-centric and integrative lighting and its potential benefits, but they consider it moderately important for the library upon receiving relevant information.

Adjustable brightness levels, lighting that adapts to natural daylight, and task-specific lighting options are identified as top priorities for future improvement.

In conclusion, Library users generally express satisfaction with the current lighting conditions in public libraries, although some issues related to discomfort and eye strain during specific weather conditions have been reported. There is a notable preference among users for lighting options that can be adjusted to individual preferences and tasks, particularly in specific library areas. While users are generally content with the lighting, there is limited awareness of the concept of human-centric and integrative lighting. However, they recognize its moderate importance for the library environment. The adjustable brightness levels, integration with natural daylight, and task-specific lighting options are areas identified for potential enhancements in the future, aligning with users' expectations for improved lighting experiences. Overall, users' feedback and preferences provide valuable insights for the ongoing improvement of lighting asset management in public libraries.

### D. SUMMARY AND CONCLUSIONS

A short summary and conclusion on qualitative insights from stakeholder interviews and questionnaires include:

1) All three Library managers express dissatisfaction with current lighting asset management practices and a lack of strategy and documentation.
2) Staff members generally find lighting conditions to be at a normal level but report discomfort in certain situations, emphasizing the need for adjustable lighting.
3) Users are satisfied with library lighting but also experience discomfort in specific conditions, with a preference for adjustable lighting.
4) Despite users' general satisfaction, they are less likely to communicate lighting concerns or seek guidance.
5) All stakeholders express limited familiarity with human-centric and integrative lighting.
6) The main areas for improvement identified across stakeholders include adjustable lighting options, daylight adaptation, and task-specific lighting.

Overall, we would benefit from better communication between library managers, staff, and users to address lighting concerns and enhance lighting asset management practices. The study highlights the need of ongoing monitoring in lighting conditions and improvements in lighting asset management through a human-centric and integrative approach to support user health and well-being in public libraries.

## V. DISCUSSION

This section synthesizes our study's findings, aligning field observations and insights from interviews A. It also discusses the implications for lighting practices in Swedish public libraries and considers broader applications B.

### A. SYNTHESIZING FIELD OBSERVATIONS AND INTERVIEW INSIGHTS

In this section, we converge our extensive research data from multiple sources – on-site observations, and insightful stakeholder interviews – to construct a holistic understanding of the present state of human-centric and integrative lighting asset management in public libraries.

Our field observations revealed a complex landscape of lighting conditions within public libraries. While some libraries demonstrated best practices in terms of LED applications, strategic use of glass elements for daylight integration, and user-adjustable lighting, others suffered from issues like glare, overbrightness, and inadequate lighting management. These findings prompted a critical reflection on the current lighting asset management practices.

Interviews with library managers, staff, and users offered valuable insights into their experiences, expectations, and challenges related to lighting. Library managers expressed dissatisfaction with current lighting asset management practices and an absence of well-documented strategies. Staff members highlighted issues like eye strain, discomfort, and the lack of a formal process for reporting lighting-related problems. Users, on the other hand, exhibited a range of experiences, with some reporting satisfaction and others facing discomfort, particularly in certain lighting conditions.

Synthesizing these diverse data sources, it becomes evident that while progress has been made in adopting energy-efficient LED technology, there is a noticeable gap in effective lighting asset management once libraries are in active use. Our findings underscore the benefit of transitioning toward human-centric and integrative lighting asset management, considering both the visual and non-visual effects of light. This synthesis provides a comprehensive foundation for subsequent discussions on the implications of our findings, opportunities for improvement.

### B. IMPLICATIONS FOR SWEDISH PUBLIC LIBRARIES AND BEYOND

The insights gained from our comprehensive study have far-reaching implications for Swedish public libraries and offer valuable lessons that extend beyond national boundaries.

Firstly, for Swedish public libraries, the need for a paradigm shift in lighting asset management is evident. The current lack of comprehensive strategies, documentation, and formal processes for lighting management poses challenges to providing optimal lighting conditions that support the well-being of both staff and users. Therefore, libraries should consider prioritizing the development of lighting asset management strategies that align with human-centric and integrative lighting principles. These strategies should encompass maintenance, optimization, and continuous improvement to ensure lighting conditions evolve in line with advancements in technology and user expectations.

In addition, libraries should explore the possibility of appointing dedicated lighting asset managers or specialists who can oversee lighting-related matters. These experts can bridge the gap between lighting technology advancements, user needs, and efficient asset management. Training and development programs could also be introduced to enhance the competence of library staff in dealing with lighting-related issues effectively.

Furthermore, libraries should seek to foster a culture of open communication and feedback regarding lighting conditions. Staff members and users should feel encouraged to report lighting-related problems promptly, with efficient mechanisms in place to address these issues. Libraries can consider implementing regular lighting assessments and root cause analyses to proactively identify and resolve challenges.

Beyond Swedish public libraries, the implications resonate with any institutions, organizations, or spaces that prioritize human-centric and integrative lighting. The call for a proactive and comprehensive approach to lighting asset management is universal. Institutions worldwide should recognize the importance of aligning lighting strategies with the well-being of occupants and the dynamic nature of their needs.

In conclusion, the implications of our study extend beyond the realm of Swedish public libraries, emphasizing the urgent need for a holistic approach to lighting asset management that prioritizes human health and well-being while adapting to evolving technological landscapes. These implications serve as a catalyst for transformation in lighting practices on a global scale.

## VI. CONCLUSION

To address the limitations identified in traditional reliability assessments of lighting sources under current standards, it is essential to adopt advanced asset management approaches that enhance awareness and provide a more comprehensive evaluation framework. This study delves into the current state of ''human-centric'' or ''integrative lighting'' in public libraries and assesses whether lighting asset management practices have kept pace with technological advancements like LEDs. It also explores stakeholder experiences and expectations related to human-centric lighting in public libraries. The research spans 20 diverse libraries in Sweden, employing fieldwork, interviews, questionnaires, and gap analysis to uncover insights into the challenges and aspirations of library lighting. The major takeaways from our study include:

### A. EXISTING CHALLENGES

Library managers, staff, and users expressed dissatisfaction with the current state of lighting asset management in libraries. Key challenges include inadequate documentation, undesirable lighting conditions, budget constraints, and a lack of familiarity with human-centric and integrative lighting principles.

### B. USER EXPERIENCE

Library users reported a range of lighting-related experiences, with a majority being satisfied with library lighting. However, issues such as eye strain and discomfort were prevalent, indicating room for improvement.

### C. AWARENESS GAP

There is a notable gap in awareness among library stakeholders regarding human-centric and integrative lighting principles and relevant standards. Education and awareness initiatives are needed to bridge this gap.

To pave the way for a brighter future in library lighting, it is imperative to focus on the following future directions:

### D. CONTINUOUS IMPROVEMENT

Libraries should embrace a culture of continuous improvement in lighting asset management. This involves developing documented strategies, conducting regular lighting assessments, and addressing issues promptly.

### E. USER-CENTRIC APPROACH

The user experience should be at the forefront of lighting decisions. Libraries should strive to create lighting environments that not only meet minimum standards but also cater to the diverse needs and preferences of users.

### F. AWARENESS AND TRAINING

Library professionals and users should receive training and awareness programs on human-centric and integrative lighting principles and relevant standards. This knowledge empowers stakeholders to make informed decisions and advocate for lighting improvements.

In conclusion, our study underscores the benefits of a paradigm shift towards human-centric and integrative lighting asset management in public libraries. By addressing the challenges and limitations identified in the current state of lighting asset management, we can create libraries that prioritize user well-being, adapt to technological advancements, and provide optimal lighting environments for diverse user needs and preferences. This endeavor requires collective commitment and collaboration from all stakeholders involved in the lighting, architecture, and library management sectors. Together, we can illuminate the path towards a brighter and more user-centric future for library lighting.

### ACKNOWLEDGMENT

The authors would like to thank Dr. Jörgen Sjödin, the Project Manager of Swedish Energy Agency, whose expert guidance has been indispensable throughout the research process. Lastly, they owe a great debt of gratitude to all the participants from the 20 public libraries, whose active involvement was crucial. The managers, staff, and regular users generously contributed their time and shared their experiences, offering a wealth of insights that have been central to their understanding of integrated lighting asset management in public libraries. Their input has been invaluable and they are truly appreciative of their collaboration.

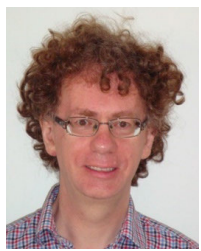

**MATH BOLLEN** (Fellow, IEEE) received the M.Sc. degree in electrical engineering and the Ph.D. degree in electric power engineering from Eindhoven University, The Netherlands, in 1985 and 1989, respectively.

Currently, he is a Chair Professor in electric power engineering with Luleå University of Technology. He is recognized as a fellow of the IEEE Power and Energy Society and a honor that underscores his significant contributions to the field.

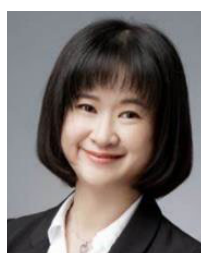

**JING (JANET) LIN** (Senior Member, IEEE) received the Ph.D. degree in management science from Nanjing University of Science and Technology, in 2008.

She currently holds dual academic positions: as a Guest Professor with the Division of Product Realization, Mälardalen University, Sweden, and as an Associate Professor with the Division of Operation and Maintenance, Luleå University of Technology, Sweden. She is also the Principal Investigator (PI) of the Project "Integrated Lighting Asset Management in Public Libraries through Digital Twins" funded by Swedish Energy Agency. Her research interests include prognostics and health management (PHM), asset management, reliability, availability, maintainability, safety, sustainability, security, supportability (RAM4S), and e-Maintenance.

Dr. Lin is also the Vice President of the IEEE Reliability Society. She played a pivotal role in establishing the IEEE Reliability Society's Sweden and Norway joint Section Chapter, in May 2021, and has been it's the Chair since its inception.

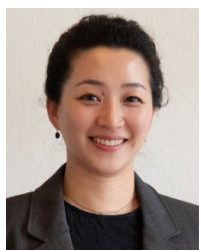

**JINGCHUN SHEN** received the M.Sc. degree in renewable energy and sustainable development from De Montfort University, U.K., in 2010, and the Ph.D. degree in solar thermal facade systems and their applications from the University of Nottingham, U.K., in 2015.

He is currently a Senior Lecturer with the School of Information and Engineering, Dalarna University, Sweden. Her areas of expertise encompass a broad spectrum of topics, including energy efficient building design, energy performance simulation and analysis of buildings (specializing in IDA ICE software), solar radiation and geometry, solar building design, introduction to bioclimatic design, sustainable green building rating systems, BIM in construction processes, and energy projects.

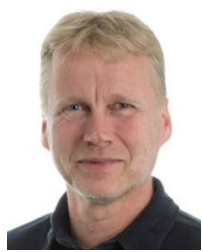

**PER OLOF HEDEKVIST** (Senior Member, IEEE) received the Ph.D. degree in photonics from the Chalmers University of Technology, Sweden, in 1998.

Currently, he is a Senior Scientist with the RISE Research Institutes of Sweden, specializing in applied metrology. His role is pivotal in developing and advancing research and innovations within the field of metrology. His areas of expertise encompass a wide range of measurements, including time and frequency, photometry and radiometry, and electricity spanning from DC to millimeter-wave. His research interests include length and positioning, mass, torque, pressure, and flow. He is particularly focused on measurements with traceability to quantum realizations, essentially encompassing all aspects of metrology that can be quantified with precision and reliability.

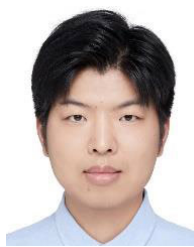

**JIAWEI XIONG** (Graduate Student Member, IEEE) received the B.Sc. and M.Sc. degrees in industrial engineering from Nanjing Tech University, Nanjing, China, in 2016 and 2019, respectively. He is currently pursuing the Ph.D. degree in management science and engineering with Nanjing University of Science and Technology, China.

He spent a year as a Visiting Researcher with Luleå University of Technology, from January to December 2023. He collaborates closely with Jiangsu Province Engineer Research Center of Quality Improvement for High-End Equipment, China. His research interests include deep learning, transfer learning, and intelligent fault diagnosis and prognosis.

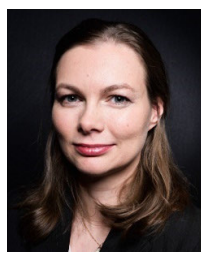

**NINA MYLLY** is currently a Distinguished Expert in lighting design. She currently holds the position of the Senior Project Leader of the Division of Measurement Technology, RISE Research Institutes of Sweden.

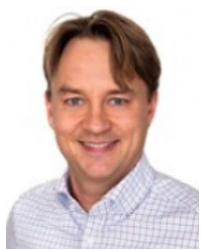

**CHRISTOFER SILFVENIUS** (Senior Member, IEEE) received the M.Sc. degree in applied physics, in 1994, and the Ph.D. degree in the design and fabrication of semiconductor lasers from the Royal Institute of Technology (KTH), Sweden, in 1999.

He is currently engaged with the SCANIA Technical Center, where he is also the Test Leader of Environmental Testing and Verification. His work primarily focuses on life length prediction for e-mobility and general electronic and electrical components.

Dr. Silfvenius is an active member of several IEEE boards. Since 2021, he has been a Board Member of IEEE Smart Lighting. He has been serving as the Board Member and the Chair for the IEEE Social Implication of Technology, since 2009, and was the Treasurer of the IEEE Sweden Section, from 2014 to 2021.